\documentclass[
reprint,
superscriptaddress,
amsmath,amssymb,
aps,
pra,
]{revtex4-1}
\usepackage{float}
\usepackage{xcolor}

\usepackage{graphicx}
\usepackage{dcolumn}
\usepackage{bm}
\usepackage[colorlinks=true, pdfstartview=FitV, linkcolor=blue, citecolor=blue, urlcolor=blue]{hyperref}
\usepackage{dsfont}

\usepackage{siunitx}
\usepackage[siunitx]{circuitikz}

\graphicspath{{figures/}}

\begin{document}

\preprint{APS/123-QED}

\title{Extracting the Lifetime of a Synthetic Two-Level System}

\author{Gabriel Margiani}
\affiliation{Laboratory for Solid State Physics, ETH Z\"{u}rich, CH-8093 Z\"urich, Switzerland.}
\author{Sebastián Guerrero}
\affiliation{Laboratory for Solid State Physics, ETH Z\"{u}rich, CH-8093 Z\"urich, Switzerland.}
\author{Toni L. Heugel}
\affiliation{Institute for Theoretical Physics, ETH Z\"{u}rich, CH-8093 Z\"urich, Switzerland.}
\author{Christian Marty}
\affiliation{Laboratory for Solid State Physics, ETH Z\"{u}rich, CH-8093 Z\"urich, Switzerland.}
\author{Raphael Pachlatko}
\affiliation{Laboratory for Solid State Physics, ETH Z\"{u}rich, CH-8093 Z\"urich, Switzerland.}
\author{Thomas Gisler}
\affiliation{Laboratory for Solid State Physics, ETH Z\"{u}rich, CH-8093 Z\"urich, Switzerland.}
\author{Gabrielle D. Vukasin}
\affiliation{Departments of Mechanical and Electrical Engineering, Stanford University, Stanford, California 94305, USA}
\author{Hyun-Keun Kwon}
\affiliation{Departments of Mechanical and Electrical Engineering, Stanford University, Stanford, California 94305, USA}
\author{James ML. Miller}
\affiliation{Departments of Mechanical and Electrical Engineering, Stanford University, Stanford, California 94305, USA}
\author{Nicholas E. Bousse}
\affiliation{Departments of Mechanical and Electrical Engineering, Stanford University, Stanford, California 94305, USA}
\author{Thomas W. Kenny}
\affiliation{Departments of Mechanical and Electrical Engineering, Stanford University, Stanford, California 94305, USA}
\author{Oded Zilberberg}
\affiliation{Department of Physics, University of Konstanz, D-78457 Konstanz, Germany.}
\author{Deividas~Sabonis}
\affiliation{Laboratory for Solid State Physics, ETH Z\"{u}rich, CH-8093 Z\"urich, Switzerland.}
\author{Alexander Eichler}
\email[Corresponding author: ]{eichlera@ethz.ch}
\affiliation{Laboratory for Solid State Physics, ETH Z\"{u}rich, CH-8093 Z\"urich, Switzerland.}

\date{\today}

\begin{abstract}
The Kerr Parametric Oscillator (KPO) is a nonlinear resonator system that is often described as a synthetic two-level system. In the presence of noise, the system switches between two states via a fluctuating trajectory in phase space, instead of following a straight path. The presence of such fluctuating trajectories makes it hard to establish a precise count, or even a useful definition, of the ``lifetime'' of the state. Addressing this issue, we compare several rate counting methods that allow to estimate a lifetime for the levels. In particular, we establish that a peak in the Allan variance of fluctuations can also be used to determine the levels' lifetime. Our work provides a basis for characterizing KPO networks for simulated annealing where an accurate determination of the state lifetime is of fundamental importance.
\end{abstract}

	\maketitle


\begin{figure}
    \includegraphics[width=\columnwidth]{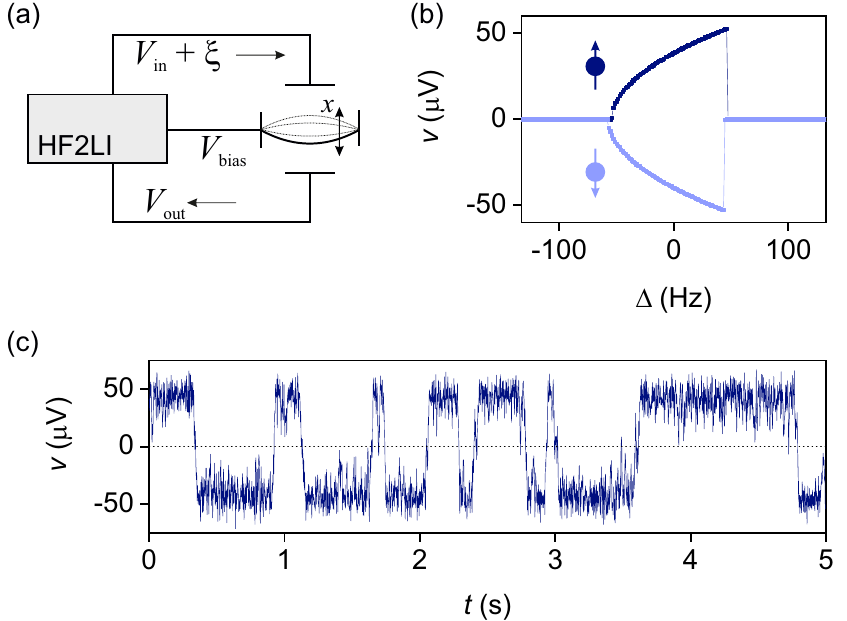}
    \caption{\textbf{Experimental setup and phase states of the KPO.} (a)~A Zurich Instruments HF2LI lock-in amplifier is used to apply a bias voltage $V_\mathrm{bias}$ to the beam, and to capacitively drive and read-out the voltage signal $V_\mathrm{out} = u \cos(\omega t) - v \sin(\omega t)$ generated by the displacement $x$ of the resonator. (b)~Measured out-of-phase response $v$ of the resonator to parametric driving as a function of detuning $\Delta = f_d-f_0$ with $V_\mathrm{in}=\SI{0.4}{\volt}$. Bright and dark dots correspond to different sweeps that showcase the amplitude-degenerate phase states of the KPO that can be interpreted as a synthetic TLS, e.g. spin-$\frac{1}{2}$ states. Each sweep contains 300 points measured within 685 seconds. (c) Switching between the phase states observed in $v$ as a function of time with $\Delta = \SI{0}{\hertz}$, $V_\mathrm{in}=\SI{0.4}{\volt}$, and $\sigma_V = \SI{1}{\volt}$. A dotted line represent the threshold between the phase states.}
    \label{fig:fig1}
\end{figure}

Synthetic two level systems (TLSs) generated in driven nonlinear resonators have recently caught a significant attention in the physics community~\cite{Gottesman_2001, Devoret_2013}. A particularly prominent example is the Kerr Parametric Oscillator (KPO, also known as parametron) \cite{Mahboob_2008, Wilson_2010, Eichler_2011_NL, eichler2018parametric, Gieseler_2012, Lin_2014, Puri_2017, Nosan_2019, Frimmer_2019, Grimm_2019, Puri_2019_PRX, Miller_2019_phase, Ryvkine_2006} whose potential energy is pumped at frequency $f_p$ close to twice its resonance frequency $f_0$, i.e. at $f_p \approx 2f_0$. If the modulation strength $\lambda$ exceeds a threshold $\lambda_\mathrm{th}$, the device responds with oscillations locked to $f_d \equiv f_p/2$ within a certain detuning range. This well-known ``period doubling'' of the response relative to the pump gives rise to two stable ``phase states'' with the same amplitude but separated by a phase difference of $\pi$. The phase states can be used to encode the two polarization states (up/down) of a classical spin. This analogy leads to the idea of using networks of coupled KPOs to build noisy intermediate-scale quantum (NISQ) machines~\cite{Preskill_2018,Albash_2018}. These machines can simulate the dynamics of mathematical problems that overwhelm traditional computers, such as the ground state of an Ising Hamiltonian~\cite{Ising_1925, Goto_2018, Rota_2019, Heim_2015, Mahboob_2016, Inagaki_2016, Goto_2016, Bello_2019, Okawachi_2020, Strinati2019}, or of other complex systems that can be mapped onto the same framework ~\cite{Lucas_2014, Nigg_2017, Inagaki_2016_Science, Goto_2019,honjo2021100}.

An important quantity for many applications of TLSs is their lifetime $\tau$ \cite{krantz2019quantum}. It is the typical time spent on a level before the interaction with an environment induces a (seemingly) spontaneous ``jump'' from one state to the other. The rates of environmental noise-induced switching have previously been investigated for different systems, such as trapped electrons \cite{lapidus},
cold atoms \cite{kkim2005}, micromechanical systems \cite{Aldridge2005,Chan_2007,Venstra2013} and analog electronic circuits \cite{Luchinsky1999}.

The situation is more subtle for a KPO. Here, the synthetic levels are formed by coherent bosonic states forming attractors in phase space. These attractors are not separated by an energy gap but by a phase gap~\cite{Frimmer_2019}. When switches occur on a slow timescale (relative to the resonator relaxation time) and follow narrow channels in phase space, the fluctuations are termed ``weak''. Such a setting allows for situations with negligible backaction where the fluctuations during a single switch can be observed. Currently, however, there exist very few studies of the fascinating physics unfolding during individual switches~\cite{Chan_2008,Mahboob_2014_2,Dykman_1998,Chan_2007}.

In this work, we study a classical micromechanical KPO and investigate its switching rates in the presence of weak fluctuations.
We invoke and compare several methods previously used to characterize the rates of charge and parity state switching in Cooper pair boxes and superconducting qubits \cite{serniak2018hot,serniakDirectDispersiveMonitoring2019}. Furthermore, we propose a method to calculate the switching rate that is based on the Allan variance of the resonator displacement \cite{van1982new}. In the final part of the paper, we compare all methods and find good agreement between several (but not all) of them.

Our KPO consists of a micro-electromechanical resonator (MEMS) in a room-temperature setup schematically shown in Fig.~\ref{fig:fig1}(a). The resonator is a doubly-clamped beam, with the length of \SI{200}{\micro\metre}, width \SI{3}{\micro\metre}, and \SI{60}{\micro\metre} in thickness with a lumped mass of 25.4 ng made from highly-doped single crystal silicon and fabricated in a wafer-scale encapsulation process \cite{yang2016unified}. Electrodes on both sides separated from the conducting beam with a gap $\approx$ \SI{1}{\micro\metre} enable capacitive driving and sensitive detection of oscillations in the presence of a bias voltage, $V_\mathrm{bias} = \SI{10}{\volt}$~\cite{Miller_2019}. We use a Zurich Instruments HF2LI lock-in amplifier to apply the driving voltage $V_\mathrm{in}$ and to measure the resonator displacement $x \propto V_\mathrm{out} = u \cos(\omega t) - v \sin(\omega t)$ with quadrature amplitudes $u$ and $v$. For convenience, we drop the proportionality factor between $x$ and $V_\mathrm{out}$ and identify in the following $x$ $\equiv$ $V_\mathrm{out}$~\cite{Nosan_2019}.

\begin{figure}
    \includegraphics[width=\columnwidth]{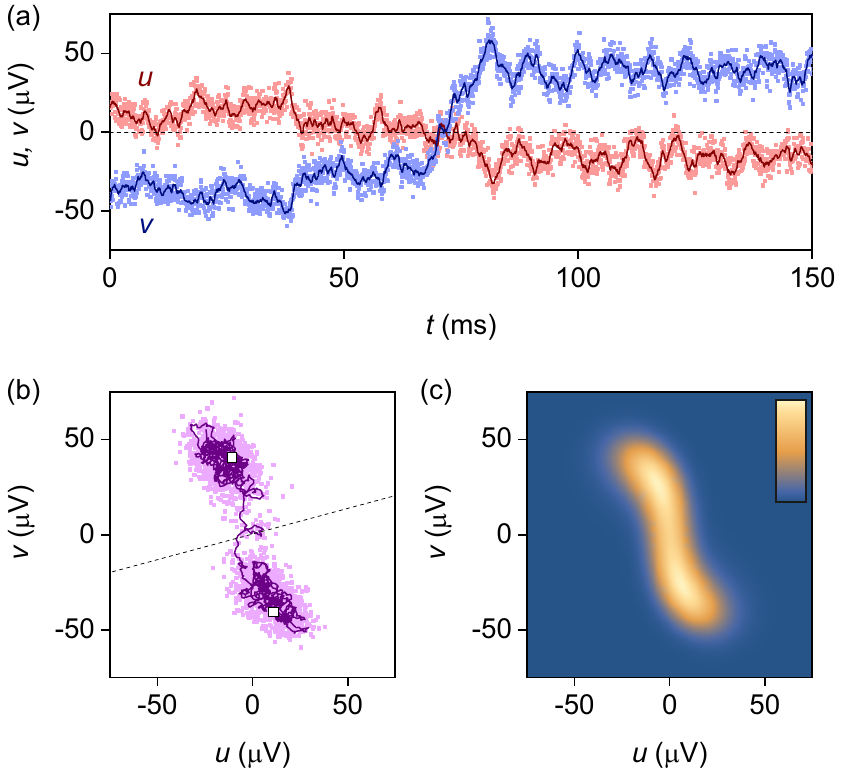}
    \caption{\textbf{Phase space representation of states and switching.} (a)~$u$ and $v$ quadratures of a single phase state switch composed of 2170 data points measured with a \SI{15}{\micro\second} integration time at 14391 samples per second. Bright dots and dark lines correspond respectively to raw data and to a 10-point moving average that allows to reduce the influence of detection noise. A dashed line indicates the threshold between the phase states. $\Delta = \SI{0}{\hertz}$, $V_\mathrm{in}=\SI{0.4}{\volt}$, and $\sigma_V = \SI{0.6}{\volt}$. (b)~Phase space representation of the data in (a). White squares indicate the attractor points measured in the absence of noise, and a dashed line indicates the threshold between the phase states. (c)~Probability density of the KPO steady state calculated with a numerical evolution of a Fokker-Planck description of the system. Driven by classical force noise, the system explores its phase space stochastically. Dark blue indicates a low probability that the KPO visits a position in phase space within a finite time, bright yellow indicates a high probability (scale not normalized).}
    \label{fig:fig2}
\end{figure}

Our mechanical resonator can be described by the nonlinear equation of motion (in units of the measured electrical signal)
\begin{multline}
	\ddot{x} + \omega_0^2\left[1-\lambda\cos\left(2\omega_d t\right)\right]x + \alpha x^3 + \gamma \dot{x} = \xi\,. \label{eq:nonlinear_EOM}
\end{multline}
Here, dots indicate time derivatives, $\omega_0/2\pi = f_0 = \SI{439.56}{\kilo\hertz}$ is the resonance frequency, $\alpha = \SI{1.47e18}{\per\volt\squared\per\second\squared}$ the coefficient of the Duffing nonlinearity, $\gamma = \omega_0/Q = \SI{770}{\hertz}$ the resonator relaxation rate, and $Q = 3580$ the quality factor of the resonator. The potential energy term ($\propto x$) is pumped with the parametric modulation depth $\lambda = 2V_\mathrm{in}/(V_\mathrm{th}Q)$ at the angular frequency $2\omega_d = 4\pi f_d$, and where $V_\mathrm{th} = \SI{320}{\milli\volt}$ is the voltage threshold for parametric oscillations for the case $f_d = f_0$ (demodulation frequency). The potential modulation arises because the electrostatic force due to $V_\mathrm{in}$ pulls the beam closer towards one electrode. The force is nonlinear, i.e., it grows stronger for small beam-electrode distances, which corresponds to a change in the overall spring constant that the beam experiences. As a consequence, the drive generates small frequency variations $\delta f_0 \propto V_\mathrm{in}$. The force term $\xi$ in Eq. \ref{eq:nonlinear_EOM} represents a fluctuating thermal bath [see Supplemental Material (SM) for details].

\begin{figure*}
    \includegraphics[width=\textwidth]{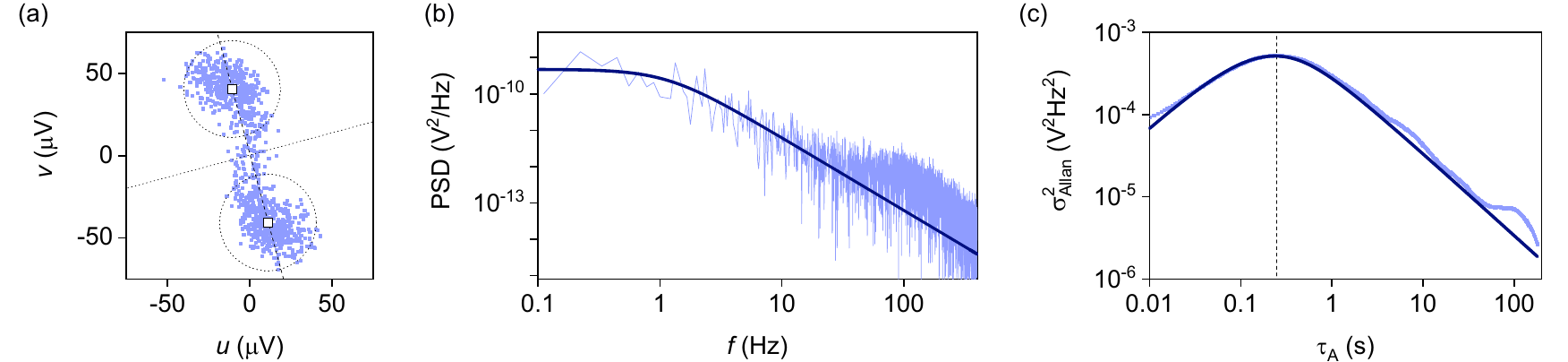}
    \caption{\textbf{Methods used to estimate the phase state lifetime.} All plots show the same \SI{15}{\minute} data set, an extract of which is shown in Fig.~\ref{fig:fig1}(c). Data was recorded at 899 samples per second with an integration time $\tau = \SI{143}{\micro\second}$ and with $\Delta = \SI{0}{\hertz}$, $V_\mathrm{in}=\SI{0.4}{\volt}$, and $\sigma_V = \SI{1}{\volt}$. (a)~Phase space representation of the two phase states and of switching between them, cf. Fig.~\ref{fig:fig2}. White squares indicate the attractors measured in the absence of noise, and the dotted line and circles indicate different threshold methods outlined in the text. The radius of the circle in this case was set to be 70 \% of the distance from the center of the circle to the origin of the coordinate system. The estimated activation rates are $\Gamma_\mathrm{line} \approx \SI{13}{\hertz} \pm \SI{0.1}{\hertz}$ and $\Gamma_\mathrm{circ} \approx \SI{4.35}{\hertz}\pm\SI{0.07}{\hertz}$, with standard deviations calculated assuming Poisson statistics of the jumps. (b)~PSD analysis of the fluctuations in terms of a telegraph noise model, cf. Eq.~\eqref{eq:PSD_telegraph}, yielding a fit result $\Gamma_\mathrm{psd} \approx \SI{3.60}{\hertz}~\pm\SI{0.01}{\hertz}$ with a fit value $F$=5.86 $\cdot 10^{-5}$ V$^{2}$. Bright and dark lines correspond to the measured data and to the fit, respectively. (c)~Allan variance of the measured fluctuations (bright), with a maximum at $\Gamma_\mathrm{Allan}$ $\approx$ $1/\tau = \SI{4.00}{\hertz}\pm\SI{0.08}{\hertz}$, where the precision is limited by the separation of points in $\tau$. A dark line is the function expected (with arbitrary vertical scaling) for pure telegraph noise with a mean switching rate of \SI{4}{\hertz}, see Eq.~\eqref{eq:Allan_telegraph}.}
    \label{fig:fig3}
\end{figure*}

Figure~\ref{fig:fig1}(b) shows the $v$-quadrature response of the resonator during two sweeps of $f_d$ from positive to negative detuning $\Delta \equiv f_d-f_0$. Close to $\Delta = \SI{50}{\hertz}$, the response jumps from $v = 0$ to $v = \pm\SI{50}{\micro\volt}$, marking a bifurcation point of the underlying nonlinear system. At the bifurcation, the resonator experiences a spontaneous $\mathds{Z}_2$ symmetry breaking, also known as a period-doubling bifurcation or a discrete time-translation breaking~\cite{Landau_Lifshitz,Heugel_2019_TC}. At this point, the resonator jumps to a positive or negative response with equal probability. The two responses belong to stable attractors ($1$ and $2$) with opposite phases, i.e., $v_1 = -v_2$ (and $u_1 = -u_2$) \cite{Strinati2019,Lifshitz_Cross,dykman1979theory, Papariello_2016}.

To study switching between the phase states of our KPO, we apply white electrical noise $\xi$ characterized by a standard deviation $\sigma_V$ (over a bandwidth of \SI{30}{\mega\hertz}) that causes the state of the resonator to fluctuate around its initial solution. If the fluctuations are large enough, they will occasionally carry the resonator across the threshold in the middle between the phase states. The resonator is then captured by the opposite attractor, corresponding to a switch of the synthetic TLS. Several such processes can be observed in Fig.~\ref{fig:fig1}(c). From this observation, it appears natural to attribute a lifetime to the inverse switching rate, $\tau = \Gamma^{-1}$. However, calculating the switching rate is not straightforward due to the fluctuating trajectory.

For a deeper understanding of the system's transient behaviour during switching events, we perform measurements with a high temporal resolution. In Fig.~\ref{fig:fig2}(a)-(b), we display a narrow time segment before, during, and after a single switch. We find many data points in the unstable zone between the two phase states. A 10-point moving average filter helps to visualize the trajectory of the system during the transition. The total switching time is roughly \SI{10}{\milli\second}, much longer than the lock-in integration time of \SI{15}{\micro\second} and the moving-average filter time of \SI{700}{\micro\second}. The measurement error of each data point is \SI{3.7}{\micro\volt}, in agreement with the measured point-to-point fluctuations, but significantly smaller than the $\sim \SI{10}{\micro\volt}$ fluctuations visible on the \SI{5}{\milli\second} scale.

Our observation depicted in Fig.~\ref{fig:fig2} demonstrates that activated switches between the phase states are not deterministic, but include prominent random elements. For instance, in the phase-space representation of the switch in Fig.~\ref{fig:fig2}(b) we can clearly see that the system performs a winding path close to the origin. In our device, the fluctuations generally have a slight preference for counter-clockwise rotations around the phase states and clockwise ones around the origin. This can be explained by the combination of the drive and the nonlinearity, which leads to an effective detuning of the fluctuations from the lock-in amplifier clock~\cite{heugel2021ghost}. In the corresponding Fokker-Planck steady-state calculation presented in Fig.~\ref{fig:fig2}(c), we therefore find a channel with a significant probability density between the phase states.

These visualizations of the fluctuating trajectories expose a fundamental problem in estimating the lifetime $\tau$: since transitions follow no straight lines, they can cross any point in phase space multiple times during a single switching event. An example of this can be observed in Fig.~\ref{fig:fig2}(b), where the averaged (dark) trajectory crosses the dotted threshold line from bottom to top, describes a clockwise winding that traverses back across the threshold, and finally crosses the line a third time before completing the switch. A simple counting algorithm will in this case register three crossing events during a single switch. In general, any counting method based on a simple threshold (such as a line) will therefore overestimate the switching number $N_\mathrm{switch}$ during the full time $T$, and therefore also $\Gamma = N_\mathrm{switch}/T$. This problem has been known since a long time.

The problem of overestimating the switching count can be reduced by defining multiple thresholds that have to be crossed in a particular order to constitute an event. In Fig.~\ref{fig:fig3}(a), we demonstrate this with the example of two circles in phase space. The count is increased by one each time a circular threshold is left and the opposing circle is entered. This method is less sensitive to small fluctuations, but it requires a subjective measure that impacts the estimated $\Gamma$, in our case the radii of the circular thresholds. Calibrating the measured switching rate $\Gamma$ as a function of the radius helps to reduce this degree of arbitrariness (see SM), but it cannot be removed entirely.

To avoid overcounting and subjective dependencies, it is desirable to extract $\Gamma$ from a method that does not require thresholds at all. Interestingly, the parity  lifetime of superconducting qubits can be determined via their charge-parity power spectral density (PSD)~\cite{Dutta_1981,riste2013millisecond,serniak2018hot}. Assuming that the switching is dominated by telegraph noise, the PSD of $v$ of our KPO can be fitted to a Lorentzian function,
\begin{equation}\label{eq:PSD_telegraph}
  \mathit{PSD}(f) = \frac{2 F^2 \tau}{4 + (2\pi f \tau)^2}\,,
\end{equation}
where the lifetime $\tau$ corresponds to the characteristic time scale between level switching events, and $F$ is a constant related to the measurement fidelity~\cite{riste2013millisecond}. In this case, the lifetime or the switching rate is related to the width of the spectral peak in the frequency domain \cite{demtroder1973laser}. To make the estimate quantitative in Fig.~\ref{fig:fig3}(b), we fit the measured displacement power spectral density with Eq.~\eqref{eq:PSD_telegraph}, yielding a third estimation for $\Gamma = 1/\tau$ lifetime.  The method can also be applied after a Fourier transform by fitting the sliding average autocorrelation with the function $AC(\Delta t) = A e^{-2\Delta t \Gamma}$ under the assumption of stationarity and ergodicity (not shown).

Crucially, the autocorrelation is intimately related to the Allan variance (see SM for the derivation). Originally invented to characterize the fidelity of clocks, the Allan variance measures the frequency fluctuations of a resonator as a function of integration time $\tau_A$. As we are interested in the time $\tau$ over which  the typical fluctuations of $u$ (or $v$) of our KPO are maximal, we apply the Allan variance formalism \cite{allan1966statistics} to the measured values,
\begin{align}
\sigma^2_\mathit{Allan}(\tau_A) = \frac{1}{2 \tau_A^2}\left< (a_{i,2} - 2a_{i,1} + a_{i,0})^2 \right>_i.
\end{align}
In this notation,
\begin{align}
a_{k,l} = \sum_{m=0}^{k + \tau_A l/\mathit{\delta t}} v(m)
\end{align}
are sums over the measured $v$ values (or $u$ values) and $\left<...\right>_i$ denotes the mean over $i$, running from $i = 1$ to $i = N-2\tau_A/\delta t$, where $N$ is the total number of data points and $\delta t$ is the sampling time. Assuming that the signal is dominated by  telegraph-like switching with lifetime $\tau$ and amplitude $B$, we obtain~\cite{van1982new}:
\begin{align}\label{eq:Allan_telegraph}
\sigma^2_\mathit{Allan}(\tau_A)=-B^2\frac{-4  \tau_A/\tau +e^{-4\tau_A/\tau }-4 e^{-2 \tau_A/\tau }+3}{4 \tau_A ^2/\tau^2}\,.
\end{align}
Hence, the maximum of $\sigma^2_\mathit{Allan}(\tau_A)$ should occur around the value $\tau_A \approx \tau = \Gamma^{-1}$. In Fig.~\ref{fig:fig3}(c), we indeed find a peak at the expected value, yielding $\Gamma \approx \SI{4}{\hertz}$. In contrast to the PSD method, the Allan variance method does not necessarily require a fitting process, as the peak can be read off directly and is easy to interpret even in the presence of noise.

\begin{figure}
    \includegraphics[width=\columnwidth]{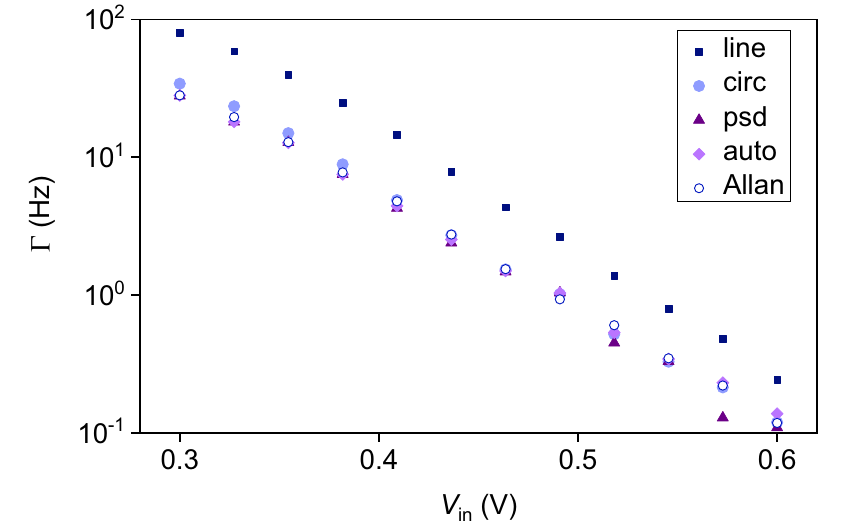}
    \caption{\textbf{Comparison of results for $\Gamma$ obtained with different rate estimation methods.} Switching rate
    as a function of parametric drive amplitude $V_\mathrm{in}$ for $\Delta = 0$ and $\sigma_V = \SI{1}{\volt}$ estimated using simple line-based thresholding (blue square), circle-based thresholding (filled circle), power spectral density of telegraph noise (triangle), autocorrelation (star) and Allan deviation (hollow circle). The radius of the circle method in this case was set to be 50 \% of the distance from the center of the circle to the origin of the coordinate system.}
    \label{fig:fig4}
\end{figure}

We compare the results of the different methods in Fig.~\ref{fig:fig4}. We find excellent agreement between four out of five of the methods for values of $\Gamma$ varying over more than two orders of magnitude. The only method that we wish to discard from this comparison is the simple line threshold approach, which consistently overestimates the count rate as expected from the discussion above. The method using two circles for thresholding overestimates $\Gamma$ slightly for $V_\mathrm{in} < \SI{0.4}{\volt}$, where the separation between the attractors is small and the ``clouds'' start to overlap significantly, cf. the example in Fig.~\ref{fig:fig3}(a). Additional comparison as a function of the noise strength $\sigma_V$ can be found in SM.

We emphasize that there is no fundamental reason why the estimators we obtain should be identical at all. The surprisingly good agreement between most of the estimators confirms that the notion of a lifetime $\tau$ is useful to characterize the switching between phase states in a KPO, where a parametric pump generates a synthetic potential landscape~\cite{Heugel_2019_TC}. This approach may be useful in other systems where multi-stable potentials in dimensions higher than one are present, such as three-dimensional protein folding or other chemical reactions. For advanced applications in the future, the resonator networks could be realized through bilinear, resonant coupling of several KPOs~\cite{Bello_2019,heugel2021ghost}  (see SM for details). For MEMS such as those studied here, bilinear coupling can be achieved in multiple ways, such as pairwise~capacitive, inductive, optical, or mechanical coupling, or indirect all-to-all coupling through a separate radio-frequency cavity.

Data availability statement: The data that support the findings of this study are available from the corresponding author upon reasonable request.

The authors have no conflicts to disclose.

\section{Supplementary Material}
The supplementary material contains theory derivations for the probability density, the Allan variance, and the autocorrelation of telegraph noise, experimental demonstrations of the dependence of the extracted switching rate on the circle threshold radius and on the noise strength, and a short summary of various coupling methods for parametric oscillators.

\section*{Acknowledgments}

Fabrication was performed in nano@Stanford labs, which are supported by the National Science Foundation (NSF) as part of the National Nanotechnology Coordinated Infrastructure under Award No. ECCS-1542152, with support from the Defense Advanced Research Projects Agency’s Precise Robust Inertial Guidance for Munitions (PRIGM) Program, managed by Ron Polcawich and Robert Lutwak. This work was further supported by the Swiss National Science Foundation through grants (CRSII5\_177198/1) and (PP00P2 190078), and by the Deutsche Forschungsgemeinschaft (DFG) - project number 449653034.

\bibliography{main}

\end{document}


\title{Supplementary Material for ``Extracting the Lifetime of a Synthetic Two-Level System"}
\author{Gabriel Margiani}
\affiliation{Laboratory for Solid State Physics, ETH Z\"{u}rich, CH-8093 Z\"urich, Switzerland.}
\author{Sebastián Guerrero}
\affiliation{Laboratory for Solid State Physics, ETH Z\"{u}rich, CH-8093 Z\"urich, Switzerland.}
\author{Toni L. Heugel}
\affiliation{Institute for Theoretical Physics, ETH Z\"{u}rich, CH-8093 Z\"urich, Switzerland.}
\author{Christian Marty}
\affiliation{Laboratory for Solid State Physics, ETH Z\"{u}rich, CH-8093 Z\"urich, Switzerland.}
\author{Raphael Pachlatko}
\affiliation{Laboratory for Solid State Physics, ETH Z\"{u}rich, CH-8093 Z\"urich, Switzerland.}
\author{Thomas Gisler}
\affiliation{Laboratory for Solid State Physics, ETH Z\"{u}rich, CH-8093 Z\"urich, Switzerland.}
\author{Gabrielle D. Vukasin}
\affiliation{Departments of Mechanical and Electrical Engineering, Stanford University, Stanford, California 94305, USA}
\author{Hyun-Keun Kwon}
\affiliation{Departments of Mechanical and Electrical Engineering, Stanford University, Stanford, California 94305, USA}
\author{James ML. Miller}
\affiliation{Departments of Mechanical and Electrical Engineering, Stanford University, Stanford, California 94305, USA}
\author{Nicholas E. Bousse}
\affiliation{Departments of Mechanical and Electrical Engineering, Stanford University, Stanford, California 94305, USA}
\author{Thomas W. Kenny}
\affiliation{Departments of Mechanical and Electrical Engineering, Stanford University, Stanford, California 94305, USA}
\author{Oded Zilberberg}
\affiliation{Department of Physics, University of Konstanz, D-78457 Konstanz, Germany.}
\author{Deividas~Sabonis}
\affiliation{Laboratory for Solid State Physics, ETH Z\"{u}rich, CH-8093 Z\"urich, Switzerland.}
\author{Alexander Eichler}
\affiliation{Laboratory for Solid State Physics, ETH Z\"{u}rich, CH-8093 Z\"urich, Switzerland.}

\maketitle

\section{Probability Density}
In this section, we briefly detail our analysis steps yielding the probability density shown in Fig.~2(c) in the main text. First, we find the deterministic equations of motion for the ``slow'' quadrature amplitudes $u$ and $v$. To this end, we use the so-called van der Pol transformation and lowest-order Krylov--Bogoliubov averaging method~\cite{krylov1947introduction,Papariello_2016, Guckenheimer_Holmes}, to replace the full time-dependent equation of motion [cf.~Eq.~(1) in the main text] by time-independent averaged equations of motion. The white noise term $\xi$ is transformed in a similar way using the method described in \cite{Khasminskii:66, Roberts:86} which yields
\begin{align}
	\label{eq:slowflow}
	\dot{u} &= -\frac{\gamma  u}{2}- \left(\frac{3 \alpha}{8 \omega_d }X^2 +\frac{\omega_0^2 - \omega_d^2}{2\omega_d}+\frac{\lambda \omega_0^2}{4 \omega_d }\right) v +\frac{1}{\sqrt{2}\omega_0}\Xi_u\,,\\
	\dot{v} &= -\frac{\gamma  v}{2} + \left(\frac{3 \alpha}{8 \omega_d }X^2 + \frac{\omega_0^2 - \omega_d^2 }{2\omega_d} -\frac{\lambda \omega_0^2}{4 \omega_d }\right)u +\frac{1}{\sqrt{2}\omega_0}\Xi_v\,,\nonumber
\end{align}
where $\omega_d=2\pi f_d$ is the angular demodulation frequency and $X^2 = u^2 + v^2$. $\Xi_u$ and $\Xi_v$ are new independent white noise processes of strength $\sigma$.
Equations~\eqref{eq:slowflow} provide a good description of the system when $\lambda$, $\gamma/\omega_0$, $(\alpha/\omega_0^2) x^2$, and $\alpha\sigma^2/\omega_0^2$ are small.

We use the stochastic differential equations~\eqref{eq:slowflow} to derive the corresponding Fokker-Planck equation~\cite{risken:96, frank:05} that describes the time evolution of the probability density $p(u,v,t)$:
\begin{multline}
		\partial_t p(u,v,t)=
		- \partial_u \left\{\frac{1}{2\omega_d} \left[ \gamma \omega_d u+ \left(\alpha \frac{3}{4} X^2 + (\omega_0^2 - \omega_d^2)+ \frac{\lambda}{2} \omega_0^2 \right)v \right) p\right]  \\
		- \partial_v \left[\frac{1}{2\omega_d} \left( \gamma \omega_d v+ \left(- \alpha \frac{3}{4} X^2 -(\omega_0^2 - \omega_d^2) + \frac{\lambda}{2} \omega_0^2 \right)u\right]  p\right\}
		+ \frac{1}{2} \left( \frac{\sigma}{\sqrt{2} \omega_d} \right)^2 (\partial_u^2 + \partial_v^2) p \,.
\end{multline}
We then solve this partial differential equation (PDE) numerically for the steady state ($\dot{p}=0$) which is reached after long times and plot the outcome for the experimental parameters in Fig.~2(c).

\section{Allan variance for a system dominated by telegraph noise}
In this section, we provide a short derivation of the Allan variance of a system subject to telegraph noise [cf.~Eq.~(5) in the main text]. The Allan variance can be calculated via the autocorrelation function. For a random telegraph signal $x$ which switches between $B$ and $-B$ the autocorrelation function is given by~\cite{lecturenotes_Yamamoto}:

\begin{equation}
\langle x(t) x(t+\tau_A) \rangle = B^2 e^{-2\tau_A/\tau}\,,
\label{eq:auto}
\end{equation}

where $1/\tau$ is the mean rate of transitions. The Allan variance is obtained by the following expectation value for the cumulative amplitudes $y(t) = \int_0^t x(t') dt'$
\begin{align}
\sigma_A^2(\tau_A) &\ueq \frac{1}{2\tau_A^2} \langle \left(y(t+2\tau_A) - 2y(t+\tau_A)+y(t)\right)^2 \rangle\\
&\ueq[\text{plug-in definition of y}]\hspace{1cm}\frac{1}{2\tau_A^2} \left\langle \left( \int_0^{t+2\tau} x(t') dt' -2 \int_0^{t+\tau_A} x(t')dt'+\int_0^t x(t') dt'\right)^2 \right\rangle\nonumber\\
&\ueq[\text{cancel terms in parenthesis}] \hspace{1cm} \frac{1}{2\tau_A^2} \left\langle \left(-\int_t^{t+\tau_A} x(t') dt'+\int_{t+\tau_A}^{t+2\tau_A} x(t') dt' \right)^2 \right\rangle\nonumber\\
&\ueq[\text{expand the square}]\hspace{0.8cm}\frac{1}{2\tau_A^2} \left\langle 2\int_0^t \int_0^t x(t') x(t'') dt'' dt'- 2\int_0^t \int_0^t x(t'+\tau_A) x(t'') dt'' dt' \right\rangle\nonumber\\
&\ueq[\text{apply Eq.~S3}]\hspace{1cm}\frac{0.8}{2\tau_A^2} 2 B^2 \int_0^{\tau_A} \int_0^{\tau_A} e^{-2|t'-t''|/\tau}-e^{-2(t'-t''+\tau_A)/\tau} dt''dt'\nonumber\\
&\ueq[\text{evaluate the integral}]\hspace{0.8cm}-B^2\frac{-4  \tau_A/\tau +e^{-4\tau_A/\tau }-4 e^{-2 \tau_A/\tau }+3}{4 \tau_A^2/\tau^2}\,.\nonumber
\end{align}
We can now look for the maximum of the obtained expression by setting the first derivative equal to 0, i.e. $d\sigma_A^2/d\tau_A \equiv 0$. Expanding the numerator in $\tau_A$ around $\tau$ up to forth order, we find that the maximum is located at $\tau_A\approx0.946\tau \approx \tau$.

\section{Investigation of circle threshold radius}

    \begin{figure*}[h]
    \includegraphics[width=\textwidth]{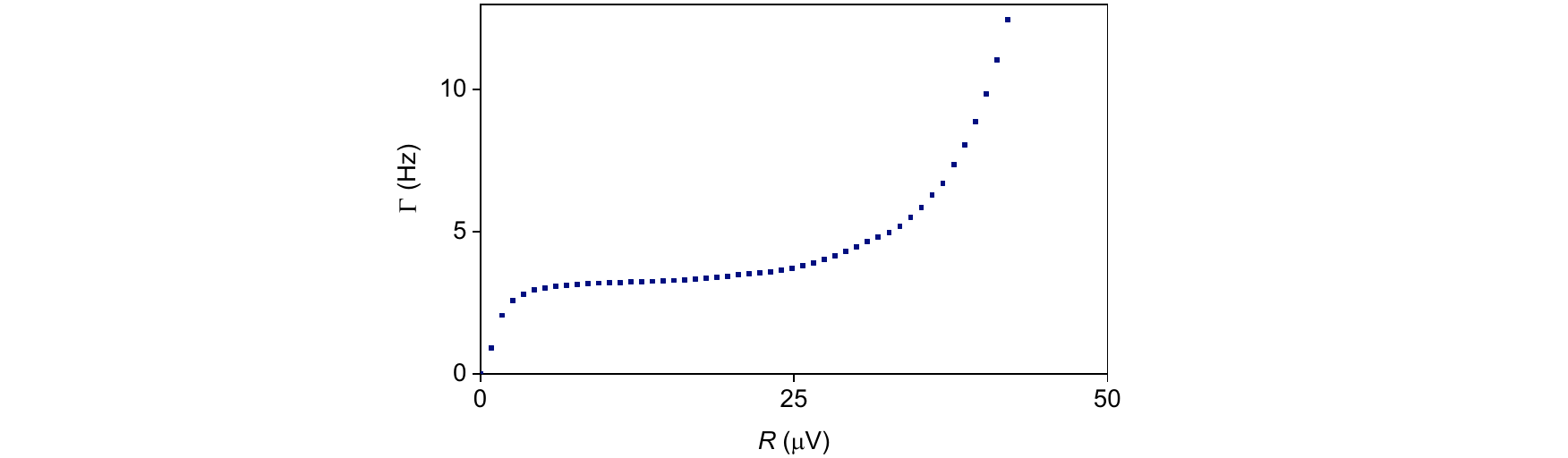} 
    \caption{\textbf{Parametron switching rate dependence on the radius of the threshold circle.}
    }
    \label{fig:SI1}
    \end{figure*}

\section{Calculation of power spectral density and autocorrelation}

The data was recorded as two quadratures $u$ and $v$ that result from the signal demodulation at half the parametric drive frequency. The quadratures were then digitized at a sampling rate of 899.46 Sa/s. The total length of the time trace was 899.94 s with a total number of  809472 recorded points. The two quadratures were then transformed into the spectral domain by using the complex ($u$+i$v$) Welch power spectral density function in Python. Afterwards, the resulting double-sided power spectral density was converted to a single sided power spectral density. As a last step, Eq. 2 in the main text was fitted to the data.

The autocorrelation $\mathit{AC}(T)$ for correlation times $T = n\mathit{\delta t}$ where $n$ is an integer and $\mathit{\delta t}$ is the sampling time, is calculated from the measured quadrature data points $u_k$ and $v_k$ in the following way. Let $a_k = u_k + iv_k$ be the complex data points, then $M_i^j=\frac{1}{j-i+1}\sum_{k=i}^{j} a_k$ is the mean of $a_k$ for $i \leq k \leq j$, and $\sigma_i^j$=$\sqrt{\frac{1}{j-i+1} \sum_{k=i}^{j}(a_k-M_i^j)\overline{(a_k-M_i^j)}}$ is the standard deviation, where $N$ the total number of data points. Here $\overline{(C)}$ denotes the complex conjugate and $\Re (C)$ the real part of a complex value $C$. The autocorrelation is then:

\begin{equation}
\textit{AC}(T = n\mathit{\delta t}) = \Re \left(\frac{1}{N-n}\frac{1}{\sigma_1^{N-n} \sigma_{n+1}^N} \sum_{i=1}^{N-n} (a_i - M_1^{N-n})\overline{(a_{i+n} - M_{n+1}^N)} \right)
\label{eq:auto_points}
\end{equation}

We fit Eq.~\eqref{eq:auto} to the resulting $\mathit{AC}(T)$ to extract the switching rate $\Gamma =1/\tau$.

\section{Switching rate as a function of $\sigma_{V}$}

Below, we show the switching rate that was estimated with all methods as a function of noise strength in a range where a two-level system exists. The parametric drive strength for this set was chosen to be $V_{in}$ = 0.436~V measured on resonance at $f$ = 0.438 MHz.

\begin{figure}[h!]
\includegraphics[width=95mm]{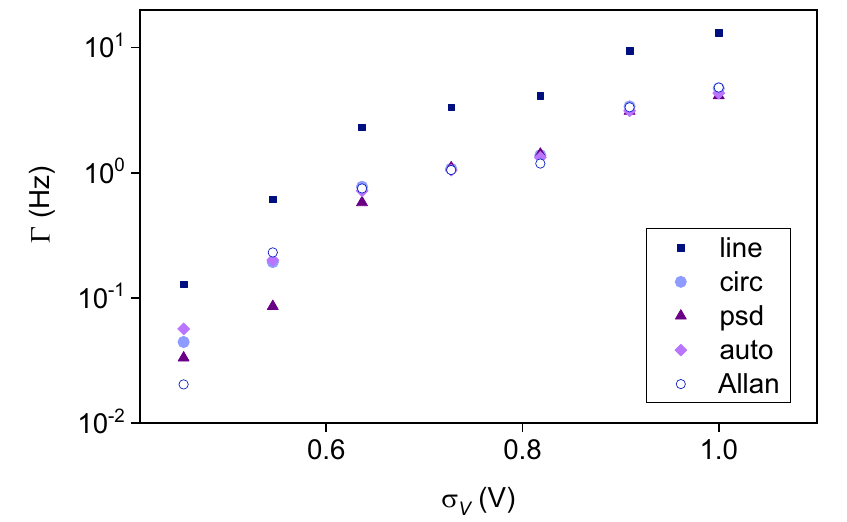}
\caption{\textbf{Comparison of results for $\Gamma$ obtained with different rate estimation methods.} Switching rate as a function of noise amplitude $\sigma_{V}$ for $\Delta = 0$ and $V_\mathrm{in} = \SI{0.436}{\volt}$ estimated using simple
average-based threshold (blue square), circle-based threshold (circles), power spectral density of telegraph noise (triangle), autocorrelation (star) and Allan deviation (hollow circle). Note that for low values of noise (roughly $\sigma_V <$ 0.6 V) the larger discrepancies between the four rate estimation methods (circle-based, PSD, autocorrelation and Allan deviation) appear due to a small switching rate (small number of switching events).}
\label{fig:SI2}
\end{figure}

We next comment on small deviations between switching rates estimated using different methods in the Fig.~\ref{fig:SI2}. In the parametron, we have two attractors in a rotating two-dimensional phase-space. One can think of the situation to be similar to a double-well potential well along one phase-space axis, alongside a single confining well along the perpendicular axis. As such, the system exhibits "fast" oscillations in the basins of phase-space as well as "slower switching" event where a phase-slip occurs and the oscillator moves from one phase state to the other. This is the reason behind the fact that the noise exhibits more involved dynamics than in a 1D double-well potential.

Our study here focuses on comparing methods by which to extract the different fluctuation sources, and specifically estimate the slow activation dynamics. To this end, we use the phenomenological  observation that the activation rates are slower than those around the attractor, and evaluate the PSD, auto-correlation, and Allan-variance based on the assumption of having solely telegraph noise. We search for clear signatures of the slow dynamics, where we can differentiate them from the fast ones.

\section{Coupling of multiple micromechanical resonators}

Resonator networks for advanced applications could be realized through bilinear, resonant coupling of several KPOs~\cite{Bello_2019,heugel2021ghost}. For micromechanical resonators such as those studied here, bilinear coupling can be achieved in multiple ways, the most common being: pairwise (I)~capacitive (II)~inductive, (III)~optical or, (IV)~mechanical coupling, or (V)~indirect all-to-all coupling through a separate radio-frequency cavity. Commonly, in the design of networks, the most complex task will be the design of multiple connections with tunable interaction strength. The following classes of coupling architectures can be envisaged:

\textit{Proximity coupling.} Exchange of photons/phonons between resonators leads to nearest-neighbor coupling. For mechanical resonators, vibrations transmitted evanescently through the substrate or electrostatic interactions through the vacuum are sources of proximity coupling. The coupling can be set via geometric design but generally not tuned in the finished device.

\textit{Connector lines.} Electrical striplines or optical waveguides that are proximity coupled to two separate resonators can act as a relay to achieve coupling between distant devices. An added advantage is that variable coupling can be obtained through changes in the transmission coefficient of the line. For instance, electrical connector lines can contain an inductive element or a tunable resonator whose transmission can be tuned.

The \textit{Lechner-Hauke-Zoller (LHZ) model.} All-to-all coupling between more than two or three devices poses a severe challenge due to the necessity of a complex network of connector lines. In the \textit{minor embedding scheme}~\cite{Choi_2008,Choi_2011}, this problem is partially mitigated by creating chains of strongly coupled KPOs that are weakly coupled to other chains, where each chain effectively corresponds to a single logic unit. The LHZ model, on the other hand, presents a more radical possibility to build an all-to-all network using individually adjustable couplings, but without the complexity of a physical connector network~\cite{Lechner_2015, Puri_2017_NC}. In the LHZ model, the pairwise orientation of network states is encoded in additional parametron devices, and there is no need for physical all-to-all connections.

\textit{Nondegenerate resonators with parametric coupling.} Many implementations of KPOs rely on degenerate parametric devices, i.e., they all have nominally identical resonance frequencies $\omega_i = \omega_j = \omega_0$ and driving frequencies at $2\omega_0$. In general, nondegenerate resonators have negligible coupling. However, three-wave mixing~\cite{Abdo_2013, Gao_2018} offers a method for sizable and tunable coupling between nondegenerate parametrons; devices with different resonance ($\omega_i \neq \omega_j$ for all pairs) and driving frequencies ($2\omega_i$, $2\omega_j$, etc.) are coupled to each other through a common medium, for instance a low-$Q$ optical cavity~\cite{Grimm_2019} that is proximity-coupled to all resonators. Modulating the cavity field at the frequency differences ($\omega_i - \omega_j$) generates parametric frequency up- and down-conversion between device pairs~\cite{Tucker_1969, Turner_2008}. The parametrons are thus in effect resonantly coupled.

\bibliography{main}